\documentclass[11pt]{article}
\usepackage{amsmath,amssymb,epsfig,cite}
\advance\hoffset by -1.1truecm
\advance\voffset by -1.0truecm
\newcommand{\be}{\begin{equation}}
\newcommand{\ee}{\end{equation}}
\newcommand{\bea}{\begin{eqnarray}}
\newcommand{\eea}{\end{eqnarray}}

\numberwithin{equation}{section}

\newcounter{appendice}

\begin{document}

\title{\begin{flushright}
 \small SU-4252-859\\IISc/CHEP/11/07
\end{flushright}
\vspace{0.25cm} Twisted Gauge and Gravity Theories on the
Groenewold-Moyal Plane} 
\author{A. P. Balachandran$^a$\footnote{bal@phy.syr.edu} ,
A. Pinzul$^b$\footnote{apinzul@fma.if.usp.br} ,
  B. A. Qureshi$^a$\footnote{bqureshi@phy.syr.edu}
  and S. Vaidya$^c$\footnote{vaidya@cts.iisc.ernet.in}\\ \\
$^a$\begin{small}Department of Physics, Syracuse University, Syracuse NY,
13244-1130, USA. \end{small} \\
$^b$\begin{small}Instituto de F\'{i}sica, Universidade de S\~{a}o
  Paulo, C.P. 66318, S\~{a}o Paulo, SP, 05315-970, Brazil. \end{small} \\
$^c$\begin{small}Centre for High Energy Physics, Indian
Institute of Science,
Bangalore, 560012, India.
\end{small}}
\date{\empty}

\maketitle
\begin{abstract}
Recent work \cite{bmpv,aschieri} indicates an approach to the
formulation of diffeomorphism invariant quantum field theories (qft's)
on the Groenewold-Moyal (GM) plane. In this approach to the qft's,
statistics gets twisted and the $S$-matrix in the non-gauge qft's
become independent of the noncommutativity parameter $\theta^{\mu
\nu}$. Here we show that the noncommutative algebra has a commutative
spacetime algebra as a substructure: the Poincar\'{e}, diffeomorphism
and gauge groups are based on this algebra in the twisted approach as
is known already from the earlier work of \cite{aschieri}. It is
natural to base covariant derivatives for gauge and gravity fields as
well on this algebra. Such an approach will in particular introduce no
additional gauge fields as compared to the commutative case and also
enable us to treat any gauge group (and not just $U(N)$). Then
classical gravity and gauge sectors are the same as those for
$\theta^{\mu \nu}=0$, but their interactions with matter fields are
sensitive to $\theta^{\mu \nu}$. We construct quantum noncommutative
gauge theories (for arbitrary gauge groups) by requiring consistency
of twisted statistics and gauge invariance. In a subsequent paper
(whose results are summarized here), the locality and Lorentz
invariance properties of the $S$-matrices of these theories will be
analyzed, and new non-trivial effects coming from noncommutativity
will be elaborated. 

This paper contains further developments of \cite{bpqv1} and a new
formulation based on its approach.
\end{abstract}

\section{Introduction}
If there is a symmetry group $G$ with elements $g$ and it acts on a
single particle Hilbert space ${\cal H}$ by the unitary representation
$g \rightarrow U(g)$, then conventionally it acts on the two-particle
Hilbert space ${\cal H} \otimes {\cal H}$  by the representation
\begin{equation}
g \rightarrow U(g) \otimes_{\mathbb C} U(g) := [U \otimes_{\mathbb C}
  U](g \otimes g).
\end{equation}
(The tensor product of vector spaces hereafter will always be over
$\mathbb{C}$.) If it acts on Hilbert spaces ${\cal H}_1$ and ${\cal
H}_2$ by representations $U_1$ and $U_2$ , then conventionally it acts
on ${\cal H}_1 \otimes {\cal H}_2$ by the representation
\begin{equation}
g \rightarrow [U_1 \otimes U_2]( g \otimes g). \label{twoparticle}
\end{equation}
The homomorphism
\begin{eqnarray}
\Delta: G &\rightarrow& G \otimes G, \nonumber \\
g &\rightarrow& \Delta(g) := g \otimes g \label{coprod}
\end{eqnarray}
underlying (\ref{twoparticle}) and (\ref{coprod}) is said to be a
coproduct on $G$. The existence of such a homomorphism is essential
for physics. For example, it is the coproduct which determines how a
diquark wavefunction transforms under color $SU(3)$, once we agree
that each quark transforms by its $\underline{3}$ representation.

Let $G^*$ be the group algebra of $G$. If $G$ admits a left- and
right-invariant measure $d \mu$, as is generally the case in physics,
and $\alpha, \beta: G \rightarrow {\mathbb C}$ are smooth compactly
supported  functions on $G$, then $G^*$ contains the generating
elements
\begin{equation}
\int d \mu (g) \alpha(g) g, \quad \int d \mu (g') \beta(g') g'
\label{groupalg}
\end{equation}
with product
\begin{equation}
\int d\mu (g) d \mu (g') \alpha(g) \beta(g') g\;g' = \int d \mu(g)
(\alpha \ast_c \beta)(g) \;g
\end{equation}
where $(\alpha \ast_c \beta)(g)$ is the convolution of $\alpha$ and
$\beta$:
\begin{equation}
(\alpha \ast_c \beta)(g) = \int d \mu(g') \alpha(g') \beta({g'}^{-1}g).
\end{equation}
It is necessary to complete the algebra generated by (\ref{groupalg})
in a suitable topology to get all of $G^*$.

The coproduct (\ref{coprod}) extends by linearity as the homomorphism
\begin{eqnarray}
\Delta: G^* &\rightarrow& G^* \otimes G^* \nonumber \\
\int d \mu (g) \alpha(g) g &\rightarrow& \int d \mu (g) \alpha(g)
\Delta(g)
\end{eqnarray}
on $G^*$. The representation $U_i$ of $G^*$ on ${\cal H}_i$,
\begin{equation}
U_i:\int d \mu (g) \alpha(g) g \rightarrow \int d \mu (g) \alpha(g)
U_i(g),
\end{equation}
induced by those of $G$, also extend to the representation $U_1
\otimes U_2$ on ${\cal H}_1 \otimes {\cal H}_2$:
\begin{equation}
U_1 \otimes U_2: \int d \mu (g) \alpha(g) g \rightarrow \int d \mu
  (g) \alpha(g) [U_1 \otimes U_2] \Delta(g).
\end{equation}

Next we outline the action of the Poincar\'{e} group, and more
generally of the diffeomorphism group, on the Groenewold-Moyal (GM)
plane ${\cal A}_\theta ({\mathbb R}^N)$. The algebra ${\cal A}_\theta
({\mathbb R}^N)$ consists of smooth functions on ${\mathbb R}^N$ with
the multiplication map
\begin{eqnarray}
m_\theta: {\cal A}_\theta ({\mathbb R}^N) \otimes {\cal A}_\theta
({\mathbb R}^N) &\rightarrow& {\cal A}_\theta ({\mathbb R}^N)\,,
\nonumber \\
\alpha \otimes \beta &\rightarrow& \alpha \;e^{\frac{i}{2}
  \overleftarrow{\partial}_\mu \theta^{\mu \nu}
  \overrightarrow{\partial}_\nu} \ \beta := \alpha \ast \beta
\label{starmult}
\end{eqnarray}
where $\theta^{\mu \nu}$ is a constant antisymmetric tensor.

Let
\begin{equation}
F_\theta = e^{\frac{i}{2} \partial_\mu \otimes \theta^{\mu \nu}
  \partial_\nu} = ``{\rm Twist \; element}'' .
\label{twistelt}
\end{equation}
Then
\begin{equation}
m_\theta (\alpha \otimes \beta) = m_0 [F_\theta \alpha \otimes
\beta] \label{starmult1}
\end{equation}
where $m_0$ is the point-wise multiplication map, also defined by
(\ref{starmult}).

Let $\phi$ be an element of the connected component of the
diffeomorphism (diffeo) group ${\cal D}_0 ({\mathbb R}^N)$ of
${\mathbb R}^N$. The connected component ${\cal P}_+^\uparrow$ of the
Poincar\'{e} group is a subgroup of ${\cal D}_0 ({\mathbb R}^N)$. For
$x \in {\mathbb R}^N$,
\begin{equation}
\phi: x \rightarrow \phi(x) \in {\mathbb R}^N.
\end{equation}
It acts on functions on ${\mathbb R}^N$ by pull-back:
\begin{equation}
\phi: \alpha \rightarrow \phi^* \alpha, \quad (\phi^* \alpha)(x) =
\alpha[\phi^{-1}(x)].
\end{equation}
The work of \cite{aschieri} based on Drinfel'd's basic paper
\cite{drinfeld} shows that ${\cal D}_0 ({\mathbb R}^N)$ acts on ${\cal
A}_\theta ({\mathbb R}^N)$ compatibly with $m_\theta$ if its coproduct
is ``twisted'' to $\Delta_\theta$ where
\begin{equation}
\Delta_\theta (\phi) = F_\theta^{-1} (\phi \otimes \phi) F_\theta.
\label{twistedcoprod}
\end{equation}
The right-hand side of (\ref{twistedcoprod}) contains polynomials in
derivatives. So it may be best to interpret $\Delta_\theta$ in terms
of ${\cal D}_0 ({\mathbb R}^N)^*$.

We denote the representation of $\phi$ on ${\cal A}_\theta ({\mathbb
  R}^N) \otimes {\cal A}_\theta ({\mathbb R}^N)$ by $\Delta_\theta
  (\phi)$ omitting symbols like $U \otimes U$ which occur in
  (\ref{twoparticle}).

The restriction to the connected component of ${\cal D}_0 ({\mathbb
  R}^N)$ is not essential. The discussion can be extended to parity
  and time-reversal \cite{abjj}.

For $\theta^{\mu \nu}=0$ and scalar bosons, statistics is imposed on
the two-particle sector by working with the symmetrized tensor product
${\cal A}_0 ({\mathbb R}^N) \otimes_s {\cal A}_0 ({\mathbb R}^N)$. It
has elements $v \otimes_s w$ where
\begin{equation}
v \otimes_s w = \frac{1}{2}[v \otimes w + w\otimes v], \quad v,w \in
{\cal A}_0 ({\mathbb R}^N).
\end{equation}
But the twisted coproduct does not preserve symmetrization
\cite{bmpv,drinfeld,oeckl},
\begin{equation}
\Delta_\theta (\phi) (v \otimes_s w) \notin {\cal A}_0 ({\mathbb R}^N)
\otimes_s {\cal A}_0 ({\mathbb R}^N)
\end{equation}
if $v$ and $w$ are not zero. We are hence obliged to twist statistics
as well. Thus let $\tau_0$ be the flip map:
\begin{equation}
\tau_0 (v \otimes w) = w \otimes v.
\end{equation}
Then
\begin{equation}
\tau_\theta :=F_\theta^{-1} \tau_0 F_\theta = F_\theta^{-2} \tau_0
\end{equation}
commutes with $\Delta_\theta (\phi)$. It is an involution,
\begin{equation}
\tau_\theta^2 = F_\theta^{-1} \tau_0^2 F_\theta = {\bf 1} \otimes
{\bf
  1} = {\rm id}
\end{equation}
and the tensor product ${\cal A}_\theta ({\mathbb R}^N)
  \otimes_{s_\theta} {\cal A}_\theta ({\mathbb R}^N)$ with twisted
  symmetrization consists of elements
\begin{equation}
v \otimes_{s_\theta} w = \frac{1}{2}[ {\rm id} + \tau_\theta](v
\otimes w). \label{twistedsymm}
\end{equation}
The space ${\cal A}_\theta ({\mathbb R}^N) \otimes_{s_\theta} {\cal
  A}_\theta ({\mathbb R}^N)$ is invariant under the twisted diffeos
  $\Delta_\theta (\phi)$.\footnote{This immediately follows from the
  observation that twisted coproduct commutes with $\tau_\theta$:
  $\Delta_\theta (g)\tau_\theta = F_\theta^{-1} \Delta_0 (g) F_\theta
  F_\theta^{-1} \tau_0 F_\theta = \tau_\theta\Delta_\theta (g) $.}

In a similar way, we can argue that the standard antisymmetrization
$({\bf 1} - \tau_0)(v \otimes w)$ is incompatible with the twisted
coproduct, and that the two-particle sector of the twisted fermions
has wavefunctions $v \otimes_{a_\theta} w$ in $\frac{1}{2}({\bf 1} -
\tau_\theta) {\cal A}_\theta ({\mathbb R}^N) \otimes {\cal A}_\theta
({\mathbb R}^N)$:
\begin{equation}
v \otimes_{a_\theta} w = \frac{1}{2}({\bf 1} - \tau_\theta)(v \otimes w)\,.
\label{twistedasymm}
\end{equation}

In standard quantum physics with $\theta^{\mu \nu}=0$, the statistics
operator $\tau_0$ is superselected: all observables commute with
$\tau_0$. Following this lead, we assume that such a superselection rule
holds also for $\theta^{\mu \nu} \neq 0$, and that all observables
commute with $\tau_\theta$.

The creation-annihilation operators of quantum fields appropriate
to (\ref{twistedsymm}) and (\ref{twistedasymm}) have been written
down before in terms of operators for $\theta^{\mu \nu}=0$
\cite{bmpv}. They will be recalled later.

In this paper, we will show that there is a representation of the
commutative algebra ${\cal A}_0 ({\mathbb R}^N)$ on ${\cal A}_\theta
({\mathbb R}^N)$. We can construct Poincar\'{e} and diffeo generators
as certain natural differential operators based on ${\cal A}_0
({\mathbb R}^N)$. Their exponentiation also gives a representation of
the associated groups. It is remarkable that acting on ${\cal
A}_\theta ({\mathbb R}^N)$, their coproduct is precisely
$\Delta_\theta$. Further considerations of this work are based on this
striking fact.

This representation of the Poincar\'{e} group on ${\cal A}_\theta
({\mathbb R}^N)$ is not new. It was first discussed by Calmet
\cite{calmet1} and analyzed further in \cite{calmet2}. Their emphasis
however differs from ours.

Section 2 constructs the commutative algebra ${\cal A}_0 ({\mathbb
  R}^N)$ which acts on ${\cal A}_\theta ({\mathbb R}^N)$. The
  Poincar\'{e} generators $M_{\mu \nu}$ and in fact vector fields $v$
  in general act on elements of ${\cal A}_\theta ({\mathbb R}^N)$ in
  the standard way for the twisted action as well. Knowing this, we
  point out that we can write any vector field $v$ (of which $M_{\mu
  \nu}$ is an example) as $v^\mu \partial_\mu$ where $v^\mu \in {\cal
  A}_0 ({\mathbb R}^N)$ and $\partial_\mu$ are the usual coordinate
  derivatives.

Section 3 contains the crucial result that the preceding actions of
$M_{\mu \nu}$ and $v$ fulfill the deformed Leibnitz rule of
\cite{aschieri} which follows from the deformed coproduct.

The deformed coproduct on diffeos is introduced for the purpose of
preserving the diffeo invariance of qft's. For $\theta^{\mu \nu}=0$,
qft's are invariant under gauge groups ${\cal G}$ based on ``global
groups'' $G$ as well, and they are fundamental for basic theory. The
Poincar\'{e} group ${\cal P}$ or the diffeo group ${\cal D} ({\mathbb
R}^N)$ acts on ${\cal G}$ and the group governing a basic theory is
the semi-direct product ${\cal G} \ltimes {\cal P}$ on Minkowski space
and ${\cal G} \ltimes {\cal D} ({\mathbb R}^N)$ for gravity plus
matter. Once we decide to preserve ${\cal P}$ or ${\cal D} ({\mathbb
R}^N)$ for $\theta^{\mu \nu} \neq 0$, it is natural to try to preserve
also ${\cal G} \ltimes {\cal P}$ and ${\cal G} \ltimes {\cal D}
({\mathbb R}^N)$. This is easily done: we just have to identify ${\cal
G}$ as the group of maps from the commutative coordinates underlying
${\cal A}_0 ({\mathbb R}^N)$ to $G$. The rest of the paper explores
the consequences of this identification. Such an identification has
been done before by \cite{aschieri}. Our development of field theories
is different from theirs.

A summary of our results is as follows. Sections 4 and 5 develop
an approach to field theories where gravity and gauge theories
without matter are identical to their commutative counterparts for
$\theta^{\mu \nu}=0$. Recall that in previous work
\cite{bpq,Pinzul:2005gx}, the independence of the $S$-matrix from
$\theta^{\mu \nu}$ was established for matter without gauge
couplings. But these dual facts about matter and connections do
not mean that all effects of $\theta^{\mu \nu}$ disappear. Pauli
principle is for example affected \cite{bmpv,cghs}. They are also
very much present in the coupling of matter and gauge fields. A
clear understanding of the latter requires an elucidation of how
gauge transformations act on matter fields, or ${\cal A}_\theta
({\mathbb R}^N)$ modules, which we do in Sections 6 and 7. In
section 8, we construct quantum noncommutative gauge theories, and
show that for a $U(1)$ gauge theory, the scattering operator is
the same as the one for usual QED. 

New effects arise for non-abelian gauge theories, with the emergence
of new types of vertices. The perturbative $S$-matrices of the above
processes are not Lorentz invariant despite all our elaborate efforts
to preserve it. (However they {\it are} unitary, consistently with
\cite{bdfp} and contrary to certain claims.) The reasons for this will
be elaborated in a subsequent paper \cite{bpqv2}, where we will discuss the
relation between locality and Lorentz invariance of the $S$-matrix
(see also \cite{gl} in this connection).

It appears that the formulation of field theories on
$\mathcal{A}_\theta (\mathbb{R}^N)$ is not unique. Thus in particular,
even though the Hopf algebras describing the diffeo and gauge groups
are identical in our work and that of
\cite{vassilevich,aschieri:gauge,kobakhidze}, the formulations of
gravity and gauge field theories are not the same. But it is possible
to describe the connection between the two. We shall briefly do so
towards the end of sections 4 and 7.

This paper is an outgrowth of our previous work \cite{bpqv1} and
develops a new formulation of gauge theories based on its ideas.

\section{The Commutative Algebra ${\cal A}_0({\mathbb R}^N)$}

The algebra ${\cal A}_\theta({\mathbb R}^N)$, regarded as a vector
space, is a module over ${\cal A}_0({\mathbb R}^N)$. We can show this
as follows.

For any $\alpha \in {\cal A}_\theta({\mathbb R}^N)$, we can define two
operators $\hat{\alpha}^{L,R}$ acting on ${\cal A}_\theta({\mathbb
  R}^N)$:
\begin{equation}
\hat{\alpha}^L \xi = \alpha * \xi, \quad \hat{\alpha}^R \xi = \xi
* \alpha \quad {\rm for} \quad \xi \in {\cal A}_\theta({\mathbb
R}^N) \ ,
\end{equation}
where $*$ is the GM product defined by Eq.(\ref{starmult}) (or,
equivalently, by Eq.(\ref{starmult1})).  The maps $ \alpha \rightarrow
\hat{\alpha}^{L,R}$ have the properties
\begin{eqnarray}
\hat{\alpha}^L \hat{\beta}^L &=& (\hat{\alpha}\hat{\beta})^L, \\
\hat{\alpha}^R \hat{\beta}^R &=& (\hat{\beta}\hat{\alpha})^R, \label{right}\\
{[}\hat{\alpha}^L, \hat{\beta}^R] &=& 0. \label{LRcommute}
\end{eqnarray}
The reversal of $\hat{\alpha}, \hat{\beta}$ on the right-hand side of
(\ref{right}) means that for position operators,
\begin{equation}
{[}\hat{x}^{\mu L}, \hat{x}^{\nu L}] = i \theta^{\mu \nu} =
-[\hat{x}^{\mu R}, \hat{x}^{\nu R}].
\end{equation}
Hence in view of (\ref{LRcommute}),
\begin{equation}
\hat{x}^{\mu c} = \frac{1}{2} \left( \hat{x}^{\mu L} + \hat{x}^{\mu R}
\right)
\end{equation}
generates a representation of the commutative algebra ${\cal
  A}_0({\mathbb R}^N)$:
\begin{equation}
{[}\hat{x}^{\mu c}, \hat{x}^{\nu c}] = 0.
\end{equation}
Let $e_p \in {\cal A}_\theta({\mathbb R}^N)$ be the exponential
function for momentum $p$:
\begin{equation}
e_p(\xi) = e^{-i p \cdot \xi}.
\end{equation}
Then
\begin{eqnarray}
\hat{x}^{\mu c} e_p (\xi) &=& \frac{1}{2} \left( x^\mu e^{\frac{i}{2}
  \overleftarrow{\partial}^\mu \theta_{\mu \nu}
  \overrightarrow{\partial}^\nu} e_p + x^\mu \leftrightarrow e_p
  \right)(\xi) \nonumber \\
&=& \xi^\mu e^{-i p \cdot \xi} \label{pointwise}
\end{eqnarray}
where (\ref{pointwise}) involves point-wise multiplication. Since any
$\alpha \in {\cal A}_\theta({\mathbb R}^N)$ has the Fourier
representation
\begin{equation}
\alpha = \int d^N p \alpha(p) e_p,
\end{equation}
it follows that
\begin{equation}
(\hat{x}^{\mu c} \alpha)(\xi) = \xi^\mu \alpha( \xi)
\end{equation}
and that $\hat{x}^{\mu c}$ generates the commutative algebra ${\cal
  A}_0({\mathbb R}^N)$ acting by point-wise multiplication on ${\cal
  A}_\theta({\mathbb R}^N)$.

This result is implicit in the work of Calmet and coworkers
\cite{calmet1,calmet2}. 

Let us express ${\rm ad}\, \hat{x}^\mu$ defined by
\begin{equation}
\hat{x}^\mu * \alpha - \alpha* \hat{x}^\mu
\end{equation}
in terms of the momentum operator $\hat{p}_\mu = -i \partial_\mu$.
This is easily done using the explicit expression for the
star-product, Eq.(\ref{starmult}):
\begin{equation}
{\rm ad} \hat{x}^\mu \alpha = x^\mu * \alpha - \alpha * x^\mu =
i\theta^{\mu\nu}\partial_\nu \alpha = -\theta^{\mu\nu}\hat{p}_\nu\
. \label{adxandp}
\end{equation}
Hence\footnote{If $x^{\mu_0}$ is a commutative coordinate in the
  centre of ${\cal A}_\theta({\mathbb R}^N)$, then $\theta^{\mu_0
  \mu}=0,$ $\forall \mu$, and $\hat{x}^{\mu_0 c}\equiv
  \hat{x}^{\mu_0}$.}
\begin{equation}
\hat{x}^{\mu c} = \hat{x}^{\mu L} - \frac{1}{2} {\rm ad}
\hat{x}^\mu = \hat{x}^{\mu L} + \frac{1}{2} \theta^{\mu \nu}
\hat{p}_\nu.
\end{equation}
This result is the starting point of the work of Calmet et al
\cite{calmet1,calmet2}.

The connected Lorentz group ${\cal L}_+^\uparrow$ acts on functions
$\alpha \in {\cal A}_\theta({\mathbb R}^N)$ in just the usual way in
our approach with the coproduct-twist:
\begin{equation}
[U(\Lambda)\alpha](x) = \alpha(\Lambda^{-1} x)
\label{globaltrans}
\end{equation}
for $\Lambda \in {\cal L}_+^\uparrow$ and $U:\Lambda \rightarrow
U(\Lambda)$ its representation on functions. Hence the generators
$M_{\mu \nu}$ of ${\cal L}_+^\uparrow$ have the representatives
\begin{equation}
M_{\mu \nu} = \hat{x}_\mu^c \hat{p}_\nu - \hat{x}_\nu^c \hat{p}_\mu, \quad
\hat{p}_\mu = -i \partial_\mu \label{Ltrans}
\end{equation}
on ${\cal A}_\theta({\mathbb R}^N)$.

Vector fields $v$ are generators of the Lie algebra of the connected
component of the diffeomorphism group acting on functions. Just as for
$M_{\mu \nu}$, which is a special vector field, we now see that $v$
can be written as
\begin{equation}
v = v^\mu (\hat{x}^c) \partial_\mu \,.
\label{vectorfield}
\end{equation}
Both (\ref{Ltrans}) and (\ref{vectorfield}) look like the familiar
expressions for $\theta^{\mu \nu}=0$. Nevertheless, their action on
${\cal A}_\theta({\mathbb R}^N)$ must involve the twisted
coproduct. The next section explains why this is so.

\section{On the Twisted Coproduct}

Let us first check the modification of the Leibnitz rule for $M_{\mu
  \nu}$. We can write, as an identity,
\begin{equation}
M_{\mu \nu} (\alpha * \beta) = (M_{\mu \nu} \alpha)*\beta +
\alpha*(M_{\mu \nu} \beta) + \frac{1}{2} \big[ ({\rm ad} \hat{x}_\mu
\alpha) * (\hat{p}_\nu \beta) - (\hat{p}_\nu \alpha)*({\rm
  ad}\hat{x}_\mu \beta) - \mu \leftrightarrow \nu \big]
\end{equation}
which on using (\ref{adxandp}) and the antisymmetry of $\theta^{\mu
  \nu}$ gives
\begin{eqnarray}
M_{\mu \nu} (\alpha * \beta) &=& (M_{\mu \nu} \alpha)* \beta + \alpha*(M_{\mu
  \nu} \beta) \nonumber \\
&-& \frac{1}{2} \big[ ((\hat{p} \cdot \theta)_\mu \alpha)*
  (\hat{p}_\nu \beta) - (\hat{p}_\nu \alpha)*((\hat{p} \cdot
  \theta)_\mu \beta) - \mu \leftrightarrow \nu \big], \label{tleibnitz} \\
(\hat{p} \cdot \theta)_\rho &:=& \hat{p}_\lambda \theta^\lambda_\rho.
\end{eqnarray}

Thus the Leibnitz rule is twisted. The twist is exactly what is
required by the coproduct $\Delta_\theta$ \cite{cknt}:
\begin{eqnarray}
\Delta_\theta(M_{\mu \nu}) &=& \Delta_0(M_{\mu \nu}) - \frac{1}{2}
\big[ (\hat{p} \cdot \theta)_\mu \otimes \hat{p}_\nu - \hat{p}_\nu
  \otimes (\hat{p} \cdot \theta)_\mu - (\mu \leftrightarrow \nu) \big] \, ,\\
\Delta_0(M_{\mu \nu}) &=& M_{\mu \nu} \otimes {\bf 1} + {\bf 1}
\otimes M_{\mu \nu} \, .
\end{eqnarray}
Thus
\begin{equation}
m_\theta[\Delta_\theta(M_{\mu \nu}) \alpha \otimes \beta] = M_{\mu
  \nu}(\alpha * \beta).
\end{equation}

The operator $M_{\mu \nu}$ is a particular vector field. What we have
seen is that it is of the form (\ref{vectorfield}). A similar argument
shows that all the ``twisted'' vector fields are of the form
(\ref{vectorfield}). The connected component of the twisted
diffeomorphism group is generated by $v$. It follows that {\it this
group is isomorphic to the connected component ${\cal D}_0({\mathbb
R}^N)$ of the untwisted diffeomorphism group}.

\section{Implications for Pure Gravity}
The implications of this observation are striking. We discuss pure
gravity first.

Consider the covariant derivative
\begin{equation}
D_\mu = \partial_\mu + \Gamma_\mu + \omega_\mu
\end{equation}
where $\Gamma_\mu$ and $\omega_\mu$ are the Levi-Civita and spin
connections respectively.

Under diffeomorphisms, it is natural to assume that $D_\mu$ transforms
in the usual way. Since the former is generated by vector fields like
(\ref{vectorfield}), the transformed $D, \Gamma$ and $\omega$ depend
on $\hat{x}^c$. It is thus natural to assume that just as in the
commutative case, $\Gamma$ and $\omega$ depend only on
$\hat{x}^c$. [But this {\it is} an assumption, as the work of
\cite{aschieri} which uses an alternative assumption shows (see
below)].

Now consider the frame fields $e_\mu^a$. Just as for $\theta^{\mu
  \nu}=0$, we can assume that they are covariantly constant,
\begin{equation}
\partial_\mu e_\nu^a + \Gamma_{\mu \nu}^\lambda * e_\lambda^a +
\omega^a_{\mu b} * e_\nu^b = 0 \;,
\label{constante}
\end{equation}
and impose also the condition
\begin{equation}
\Gamma_{\mu \nu}^\lambda = \Gamma_{\nu \mu}^\lambda
\label{torsionfree}
\end{equation}
to eliminate torsion. Then (\ref{constante}) can be treated just as
for $\theta^{\mu \nu}=0$ if we assume that $e_\mu^a$ depends only on
$\hat{x}^{\mu c}$. In that case, the $*$'s in (\ref{constante}) can be
erased and $\omega$ can be expressed as
\begin{equation}
\omega^a_{\mu b} = (\partial_\mu e_\nu^a) e^\nu_b + \Gamma_{\mu
  \nu}^\lambda e_\lambda^a e^\nu_b.
\label{witogamma}
\end{equation}

We remark that in (\ref{constante}), we have the natural freedom to
reverse the order of the factors in the last two terms. This ordering
ambiguity has no effect for this solution available in our approach,
but can be important in other approaches.

We have not studied the possibility of other solutions for
(\ref{torsionfree}). Perhaps they exist, with $e_\mu^a$ depending on
both $\hat{x}^{\mu c}$ and $\hat{x}^{\mu L}$, but (\ref{witogamma}) is
satisfactory and we accept it.

Thus the gravity sector in our approach is based on the commutative
coordinate and its algebra is isomorphic (under suitable assumptions)
to ${\cal A}_0({\mathbb R}^N)$. Hence the gravity sector is based on
standard differential geometry. As ${\cal A}_0({\mathbb R}^N)$ admits
the usual integration, the dynamics in the gravity sector can be
described in the manner appropriate for $\theta^{\mu \nu}=0$.

In the formulation of \cite{aschieri}, the covariant derivative
$D_\mu^\ast$ acts with a $\ast$-product. In their formulation if we
use instead
\begin{equation} 
\mathcal{D}_\mu^\ast\ = \ D_\mu^\ast
e^{-\frac{i}{2}ad\overleftarrow{\partial}_\lambda \theta^{\lambda\rho}
\overrightarrow{ \partial}_\rho}
\end{equation} 
where
\begin{equation} 
D^\ast_\mu\ ad\overleftarrow{\partial}_\lambda\ := \ [\partial_\lambda,
D^\ast_\mu],
\end{equation} 
as covariant derivative, then
\begin{equation}
\mathcal{D}_\mu^\ast \ast\alpha = D_\mu^\ast\alpha
\end{equation}
where there is no $\ast$ on the right hand side. Hence
$\mathcal{D}_\mu^\ast$ is our covariant derivative described in their
formalism. Both the approaches seem consistent, differing only in the
choice of covariant derivative.

\section{Implications for Gauge Fields}

Gauge fields $A_\lambda$ transform as one-forms under
diffeomorphisms for $\theta^{\mu \nu}=0$. For $\theta^{\mu \nu}
\neq 0$, the vector fields $v^\mu$ generating diffeomorphisms
depend on $\hat{x}^c$. If an infinitesimal diffeomorphism acts on
$A_\lambda$ in a {\it conventional} way for $\theta^{\mu \nu} \neq 0$
and $A_\lambda$ and its variation $\delta A_\lambda$ are to depend
on just one combination of noncommutative coordinates, then
$A_\lambda$ can depend only on $\hat{x}^c$. This leads to the
conclusion that gauge fields are independent of $\theta^{\mu \nu}$
and are not affected by noncommutativity.

Such an inference is reasonable for another reason as well. Twisted
coproducts for diffeomorphisms are introduced to maintain them as
symmetries in gravity. But for $\theta^{\mu \nu}=0$, with gravity and
gauge fields present, the group of importance is not just ${\cal D}_0
({\mathbb R}^N)$, but its semi-direct product ${\cal G} \ltimes {\cal
D}_0 ({\mathbb R}^N)$. Once we decide to maintain ${\cal D}_0
({\mathbb R}^N)$ as a symmetry group for $\theta^{\mu \nu} \neq 0$, it
is natural to go the whole way and preserve ${\cal G} \ltimes {\cal
D}_0 ({\mathbb R}^N)$ for $\theta^{\mu \nu} \neq 0$. But elements of
${\cal D}_0 ({\mathbb R}^N)$ perform diffeomorphisms, so then we
should require that elements of ${\cal G}$ are constructed from the
elements of the algebra generated by $\hat{x}^c$. That would then say
that the abstract group ${\cal G}$ is independent of $\theta^{\mu
\nu}$.

But in our approach $D=d +A$ transforms under $g \in {\cal G}$
according to $D \rightarrow g D g^{-1}$. So if $A$ and its gauge
transform depend on just one coordinate operator, that operator is
$\hat{x}^c$.

If the focus is just on the Poincar\'{e} group, the above argument is
still valid on substituting this group for ${\cal D}_0 ({\mathbb
R}^N)$, provided $N \geq 3$. The case $N=2$ is special, since the
Poincar\'{e} group (in fact the volume preserving diffeomorphism
group) with the coproduct $\Delta_0$ is an automorphism of ${\cal
A}_\theta ({\mathbb R}^2)$.

The conclusion of the last two sections is that gravity and gauge
sectors are unaffected by noncommutativity.

In the standard approach to noncommutative gauge groups
\cite{cpst,bkv}, where covariant derivatives act with the $*$-product,
it is possible to treat only particular representations of $U(N)$
gauge groups or use enveloping algebras \cite{Jurco:2000ja} or deal
with the Seiberg-Witten map \cite{Seiberg:1999vs}. (But see Chaichian
et. al. \cite{cpst}). There is no such limitation now where the gauge
group is just that for $\theta^{\mu \nu}=0$.

In quantum Hall effect, the algebra of observables is ${\cal A}_\theta
  ({\mathbb R}^2) \otimes {\cal A}_\theta ({\mathbb R}^2)$. In a
  particular formulation of that system, covariant derivatives of the
  $U(1)$-gauge fields of electromagnetism do act in the manner we
  assume, and not with a $\ast$-product \cite{bgk}.

\section{Gauge Transformations and $*$-Products}

The Poincar\'{e} group was built up from $\hat{x}^c$, and not in any
other manner, but still its action preserves the $*$-product. We can
ask if gauge transformations based on $\hat{x}^c$ also preserve the
$*$-product.

\subsection{How the Gauge Group acts on ${\cal A}_\theta({\mathbb
    R}^N)$-Modules}

But this question needs clarification. Fields which transform
non-trivially under ${\cal G}$ or even the underlying ``global'' Lie
group $G$ are not elements of the algebra ${\cal A}_\theta ({\mathbb
R}^N)$. Rather they are modules over ${\cal A}_\theta ({\mathbb
R}^N)$. If a $d$-dimensional representation of $G$ is involved, they
can be elements of ${\cal A}_\theta ({\mathbb R}^N) \otimes {\mathbb
C}^d$. They may also be elements of non-trivial projective modules
(see for example Chapter 5 of\cite{bkv}). We focus on ${\cal A}_\theta
({\mathbb R}^N) \otimes {\mathbb C}^d$ for simplicity.

There are two separate matters we have to resolve about these
modules. First, we must understand the action of gauge transformations
on these modules and show their compatibility with the $*$-product. We
argue that we can accomplish such compatibility if the gauge group
also has a twisted coproduct. This twist is in fact needed to maintain
the semi-direct product structure of ${\cal G} \ltimes {\cal D}_0
({\mathbb R}^N)$ at the level of coproducts. Secondly we must show how
to form gauge scalars out of elements of ${\cal A}_\theta ({\mathbb
R}^N) \otimes {\mathbb C}^d$ and their adjoints compatibly with the
above twisted coproduct. This is an essential step in constructing
observables like the Hamiltonian. Below we describe how to accomplish
both these tasks successfully. Certain familiar structures available
for $\theta^{\mu \nu}=0$ are not available for $\theta^{\mu \nu} \neq
0$. Gauge theories for $\theta^{\mu \nu}=0$ and $\theta^{\mu \nu} \neq
0$ are thus structurally different.The section finally briefly
discusses these differences.

The results presented in the section are not new and are due to
\cite{aschieri}. So this section can be treated as a review.

Elements $\xi$ of
\begin{equation}
{\cal A}_\theta ({\mathbb R}^N)^d := {\cal A}_\theta ({\mathbb R}^N)
\otimes {\mathbb C}^d
\end{equation}
are $d$-dimensional vectors $(\xi_1, \xi_2, \cdots, \xi_d)$ where
$\xi_i \in {\cal A}_\theta ({\mathbb R}^N)$. There is an action
\begin{eqnarray}
m_\theta: {\cal A}_\theta ({\mathbb R}^N)^d \otimes {\cal A}_\theta
({\mathbb R}^N) &\rightarrow& {\cal A}_\theta ({\mathbb R}^N)^d, \\
\xi \otimes \alpha &\rightarrow& m_\theta(\xi \otimes
\alpha):=\xi*\alpha, \quad \alpha \in {\cal A}_\theta ({\mathbb R}^N),
\label{moduleprop1} \\
(\xi*\alpha)_i &\equiv& \xi_i * \alpha,
\end{eqnarray}
expressing the module property of ${\cal A}_\theta ({\mathbb
  R}^N)^d$. We treat it as a right-module for convenience.

Now if $g(\hat{x}^c)$ is a $d \times d$ matrix $\in {\cal G}$, it
transforms $\xi * \alpha$ to $g(\hat{x}^c)(\xi * \alpha)$ where
\begin{equation}
[g(\hat{x}^c)(\xi * \alpha)]_i (x) = g_{ij}(x) (\xi * \alpha)_j (x).
\label{ncgauge}
\end{equation}
But when $g_{ij}(x)$ is not a constant,
\begin{equation}
{\rm RHS \;\;of \;\; (\ref{ncgauge})} \neq (g_{ij}(x) \xi_j)* \alpha\, .
\end{equation}
Infinitesimally, for
\begin{equation}
g(\hat{x}^c) \simeq {\bf 1} + i \Lambda(\hat{x}^c),
\end{equation}
we find from (\ref{ncgauge}) that
\begin{equation}
\Lambda(\hat{x}^c)_{ij} [\xi * \alpha]_j = (\Lambda_{ij} \xi_j)*
\alpha(x) + {\rm extra}\;\;{\rm terms}
\end{equation}
which is very much like the deformed Leibnitz rule (\ref{tleibnitz}).

Let $\epsilon$ be the ``counit'', the trivial representation of ${\cal
G}$:
\begin{equation}
\epsilon(g(\hat{x}_c))={\bf 1}.
\end{equation}
Then
\begin{equation}
F_\theta^{-1} (id \otimes \epsilon)[g(\hat{x}^c) \otimes g(\hat{x}^c)]
F_\theta
\end{equation}
acts on ${\cal A}_\theta ({\mathbb R}^N)^d \otimes {\cal A}_\theta
({\mathbb R}^N)$ according to
\begin{equation}
\xi \otimes \alpha \rightarrow F_\theta^{-1} [g(\hat{x}^c) \otimes
  {\bf 1}]F_\theta (\xi \otimes \alpha),
\end{equation}
which under $m_\theta$ becomes
\begin{equation}
g(\hat{x}^c)[\xi * \alpha]
\end{equation}
which in component form is (\ref{ncgauge}).

We thus see that just as the coproduct on diffeos, the twisted
coproduct on ${\cal G}$,
\begin{equation}
\Delta_\theta (g(\hat{x}^c) = F_\theta^{-1} [g(\hat{x}^c) \otimes
  g(\hat{x}^c)] F_\theta \;,
\label{twistedgauge}
\end{equation}
is compatible with the $*$-multiplication in (\ref{moduleprop1}).

We need this twisted coproduct in any case in order that
$\Delta_\theta(\phi)$ [cf. (\ref{twistedcoprod})] acts on
$\Delta_\theta (g(\hat{x}^c)$ compatibly with the semi-direct product
structure ${\cal G} \ltimes {\cal D}_0 ({\mathbb R}^N)$.

\subsection{On Gauge Scalars}

If $\eta \in {\cal A}_\theta ({\mathbb R}^N)^d$, and it transforms
under $g(\hat{x}^c) \in {\cal G}$ according to
\begin{equation}
\eta(x) \rightarrow [g(\hat{x}^c) \eta](x) = (g_{ij}(\hat{x}^c)
\eta_j)(x) = g_{ij}(x) \eta_j(x),
\label{sectionvb}
\end{equation}
then $\eta^\dagger$ necessarily transforms as
\begin{equation}
\eta^\dagger \rightarrow (g \eta)^\dagger, \quad \eta^\dagger_i (x)
\rightarrow \eta^\dagger_j (x) g^*_{ji}(\hat{x}^c) = \eta^\dagger_j
(x) g^*_{ji}(x) \ .
\label{adjsection}
\end{equation}
If $\xi$ and $\xi^\dagger$ form another such pair, consider $\sum_i
\xi^*_i * \eta_i \equiv \xi^\dagger * \eta$. It is not invariant if
$\xi^\dagger$ and $\eta$ are naively transformed as in
(\ref{sectionvb}) and (\ref{adjsection}). But we want its invariance
only for the twisted coproduct (\ref{twistedgauge}). To check if this is
so, we define the ``multiplication'' map
\begin{equation}
\delta_\theta : \xi^\dagger \otimes \eta \rightarrow \xi^\dagger *
\eta = \delta_0 (F_\theta \xi^\dagger \otimes \eta).
\end{equation}

The representation of $g(\hat{x}^c)$ on $\xi^\dagger$ can be
denoted by $\bar{id}$, that on $\eta$ being $id$. Then
\begin{equation}
\delta_\theta [F_\theta^{-1} (\bar{id} \otimes id) (g(\hat{x}^c) \otimes
  g(\hat{x}^c)) F_\theta \xi^\dagger \otimes \eta] = \xi^\dagger * \eta
\end{equation}
showing its invariance.

\subsection{Transformations of Composite Operators}

For $\theta^{\mu \nu}=0$, if $\psi$ and $\chi$ transform by a gauge
group ${\cal G}$ as dictated by the representations $\rho$ and
$\sigma$ of its global group $G$,
\begin{eqnarray}
\psi(x) &\rightarrow& \rho[g(x)] \psi(x), \quad \chi(x) \rightarrow
  \sigma[g(x)]\chi(x), \quad g \in G, \; g(x) \in {\cal G}, \\
\psi_i (x) &\rightarrow& \rho[g(x)]_{ij} \psi_j(x), \quad
\chi_\alpha (x)
  \rightarrow \sigma[g(x)]_{\alpha \beta} \chi_\beta (x),
\end{eqnarray}
we can consistently assign a transformation law under ${\cal G}$ to
$\psi\otimes ' \chi$,
\begin{equation}
(\psi \otimes' \chi)_{i \alpha} (x,x) \equiv \psi_i (x) \chi_\alpha
(x).
\end{equation}
It is dictated by the representation $\rho \otimes \sigma$ of $G$:
\begin{eqnarray}
[\psi \otimes' \chi]_{i \alpha}(x) &\rightarrow& \rho[g(x)]_{ii'}
\psi_{i'}(x) \sigma[g(x)]_{\alpha \alpha'} \chi_{\alpha'}(x)
\label{ot1}\\
&=& \rho[g(x)]_{ii'} \sigma[g(x)]_{\alpha \alpha'} \phi_{i'}(x)
\chi_{\alpha'}(x). \label{ot2}
\end{eqnarray}
In the passage from (\ref{ot1}) to (\ref{ot2}), commutativity of
spacetime algebra has been used.

We use equations such as (\ref{ot2}) in forming gauge invariants such
as the Yukawa term in the Lagrangian density. It is used as well to
form covariant composite local fields such as a color $\bar{3}$
composite of two quark fields.

$\psi \otimes' \chi$ is not the tensor product $\psi \otimes \chi$ of
$\psi$ and $\chi$. $\psi \otimes \chi$ is a function on ${\mathbb R}^N
\otimes {\mathbb R}^N$ with value $\psi (x) \otimes \chi(y)$ at
$(x,y)$ whereas $\psi \otimes' \chi$ is a function of just $(x,x)$,
that is, $x$.

We can interpret this restriction in two different ways: \\
a) $\psi \otimes' \chi$ is the restriction of $\psi \otimes \chi$ to
the diagonals $(x,x)$. \\
b) $(\psi \otimes' \chi)_{i \alpha}$ is the product in the algebra,
being the $\ast$-product $\psi_i \ast \chi_\alpha$ if $\theta^{\mu\nu}
\neq 0$.

For $\theta^{\mu \nu} \neq 0$ these two interpretations have
different implications, although for $\theta^{\mu \nu} =0$, they
coincide. Only b) is suitable for $\theta^{\mu \nu} \neq 0$ as we will now
argue.

{\it a) Restriction to diagonals:} For $\theta^{\mu \nu} \neq 0$,
${\cal G}$ acts on $\psi \otimes \chi$ by the coproduct
(\ref{twistedgauge}). But this action is not compatible with the
restriction to $(x,x)$. We can see this in the following way:
\begin{eqnarray}
&&(\psi \otimes \chi)(x,y) \rightarrow (\rho \otimes \sigma)
    F_\theta^{-1} [g(\hat{x}^c) \otimes g(\hat{x}^c)]F_\theta (\psi
    \otimes \chi)(x,y),\\ 
&=& \exp \left( -\frac{i}{2} \frac{\partial}{\partial x} \wedge
    \frac{\partial}{\partial y}\right) \left(\rho[g(\hat{x}^c)]\otimes
    \sigma[g(\hat{y}^c)]\right) \exp \left( \frac{i}{2}
    \frac{\partial}{\partial x} \wedge \frac{\partial}{\partial
    y}\right) (\psi \otimes \chi)(x,y)
\end{eqnarray}
where $\frac{\partial}{\partial x} \wedge \frac{\partial}{\partial
y}:=\theta^{\mu\nu}\frac{\partial}{\partial x^\mu} \otimes
\frac{\partial}{\partial y^\nu}$.

This is complicated at $x=y$ and involves derivatives of gauge
transformations. Its components do not reduce to the analog
\begin{equation}
\left(\rho[g(\hat{x}^c)]_{ii'} \otimes \sigma[g(\hat{x}^c)]_{\alpha
  \alpha'}\right) (\psi_{i'} \otimes \chi_{\alpha'})(x,x)
\label{analog}
\end{equation}
of (\ref{ot2}). So $(\psi \otimes' \chi)(x,x)$ has no simple
transformation law under ${\cal G}$.

{\it b) The $*$-product:} In this case the transformation of $\psi
\otimes' \chi$ is given by
\begin{equation}
m_\theta \{ (\rho \otimes \sigma) \Delta_\theta (g) \psi \otimes \chi
\}_{i \alpha},
\label{compositestar}
\end{equation}
where
\begin{equation}
(\xi \otimes \eta)(x,y) := \xi_i (x) \eta(y)_\alpha
\end{equation}
To simplify (\ref{compositestar}), we write the twist element
$F_\theta$ (defined in (\ref{twistelt})) in
the Sweedler notation:
\begin{equation}
F_\theta = e^{\frac{i}{2}\partial_\mu \otimes \theta^{\mu \nu}
  \partial_\nu} = \sum_{n=0}^\infty \frac{(i/2)^n}{n!}\theta^{\mu_1
  \nu_1}\ldots \theta^{\mu_n \nu_n} \partial_{\mu_1} \ldots
  \partial_{\mu_n} \otimes \partial_{\nu_1} \ldots \partial_{\nu_n}
  \equiv \sum_\gamma f^{(1)\gamma} \otimes f^{(2)}_\gamma.
\end{equation}
Then (\ref{compositestar}) is
\begin{equation}
m_\theta \{ (\rho \otimes \sigma) \Delta_\theta (g) \psi \otimes
\chi \}_{i \alpha} = \sum_{\gamma,j,\beta}
\{\rho(g(\hat{x}^c))_{ij} f^{(1)\gamma} \psi_j \} \{
\sigma(g(\hat{x}^c))_{\alpha \beta} f^{(2)}_\gamma \,  \chi_\beta
(x) \} \, . \label{starprod}
\end{equation}
As there is no $*$ in (\ref{starprod}) and the gauge transformations are
as for $\theta^{\mu \nu} = 0$,
\begin{eqnarray}
m_\theta \{ (\rho \otimes \sigma) \Delta_\theta (g) \psi \otimes
\chi \}_{i \alpha} &=&\sum_{\gamma,j,\beta}
\{\rho(g(\hat{x}^c))_{ij} f^{(1)\gamma} \psi_j \} \{
\sigma(g(\hat{x}^c))_{\alpha \beta} f^{(2)}_\gamma \, \chi_\beta
(x) \}\, .\\
&=&\rho[g(\hat{x}^c)]_{ij} \sigma[g(\hat{x}^c)]_{\alpha \beta}
(\psi_{j}* \chi_{\beta})(x)\ .\nonumber
\end{eqnarray}
This is similar to (\ref{ot2}) so that composite gauge transformations
can be consistently defined.

\section{On Covariant Derivatives of Quantum Fields}

Suppose we have a charged scalar field $\phi$,
\begin{equation}
\phi(x) = \int d\mu(p) (a_p e^{-i p \cdot x} + b^\dagger(p) e^{i p
  \cdot x})
\end{equation}
that obeys twisted statistics in Fock space:
\begin{eqnarray}
a(p) a(q) &=& e^{i p \wedge q} a(q) a(p) , \\
a(p) a^\dagger (q) &=& e^{-i p \wedge q} a^\dagger (q) a(p) + 2p_0
\delta^{(3)}(p-q), \\
b(p) b(q) &=& e^{i p \wedge q} b(q) b(p) , \\
b(p) b^\dagger (q) &=& e^{-i p \wedge q} b^\dagger (q) b(p) + 2p_0
\delta^{(3)}(p-q), \\
a(p) b(q) &=& e^{ip \wedge q} b(q) a(p), \\
a(p) b^\dagger (q) &=& e^{-ip \wedge q} b^\dagger(q) a(p), \\
a^\dagger (p) b^\dagger (q) &=& e^{ip \wedge q} b^\dagger(q) a^\dagger(p).
\end{eqnarray}
As shown elsewhere \cite{replyto,bpq}, these relations are direct
consequences of the twisted statistics of the multiparticle states
discussed in Section 1.

Now the twisted operators $a(p), a^\dagger(p), b(p)$ and $b^\dagger(p)$
can be realized in terms of untwisted Fock space operators $c(p),d(p)$
as
\begin{eqnarray}
a(p) &=& c(p) e^{-\frac{i}{2} p \wedge P}, \quad a^\dagger (p) = c^\dagger (q)
e^{\frac{i}{2} p \wedge P}, {\rm where} \label{atoc} \\
P_\mu &=& \int d\mu(q) q_\mu [a^\dagger(q) a(q) + b^\dagger (q) b(q)] =
{\rm the \,\,total \,\, momentum\,\, operator} .
\end{eqnarray}
Then $\phi(x)$ may be written in terms of the ordinary or commutative
fields $\phi_c$ as
\begin{equation}
\phi(x) = \phi_c e^{\frac{1}{2}\overleftarrow{\partial} \wedge P}(x)\,.
\label{ncfieldmap}
\end{equation}
If $\phi'$ is another such quantum field, $\phi'(x) = \phi'_c
e^{\frac{1}{2}\overleftarrow{\partial} \wedge P}(x)$, then
\begin{equation}
(\phi * \phi')(x) = (\phi_c
  \phi'_c)e^{\frac{1}{2}\overleftarrow{\partial} \wedge P}(x)  
\end{equation}
(\ref{atoc}, \ref{ncfieldmap}) are the ``dressing transformations'' of
Grosse, Zamolodchikov and Faddeev \cite{grosse,zz,faddeev}.

To define the desirable properties of covariant derivatives $D_\mu$,
let us first look at ways of multiplying the field $\phi$ by function
$\alpha_c \in {\cal A}_0({\mathbb R}^4)$. There are two possibilities:
\begin{eqnarray}
\phi &\rightarrow (\phi_c \alpha_c) e^{\frac{1}{2}
  \overleftarrow{\partial} \wedge P} \equiv T_0 (\alpha_c)\phi
  \label{commrep}\\
\phi &\rightarrow (\phi_c \ast_\theta \alpha_c) e^{\frac{1}{2}
  \overleftarrow{\partial} \wedge P} \equiv T_\theta (\alpha_c)\phi
  \label{starrep}
\end{eqnarray}

In (\ref{commrep}), $T_0$ gives a representation of the commutative
algebra of functions, whereas $T_\theta$ in (\ref{starrep}) gives that
of the $\ast$-algebra.

For $D_\mu$ to qualify as the covariant derivative of a quantum field
associated with ${\cal A}_0({\mathbb R}^4)$, we require of it that
\begin{description}
\item[1]
\begin{equation}
D_\mu (T_0(\alpha_c) \phi) = T_0(\alpha_c) (D_\mu \phi) +
   T_0(\partial_\mu \alpha_c) \phi \,. \label{commgauge}
\end{equation}
\item[2] $D_\mu$ preserve statistics.
\item[3] $D_\mu$ preserve Poincar\'{e} and gauge invariance.
\end{description}
The requirement (\ref{commgauge}) reflects the fact that $D_\mu$ is
associated with the commutative algebra ${\cal A}_0({\mathbb
  R}^4)$. 

There are two immediate choices for $D_\mu \phi$:
\begin{eqnarray}
&1.& D_\mu \phi = ((D_\mu)_c \phi_c)e^{\frac{1}{2}
   \overleftarrow{\partial} \wedge P} ,\label{rightchoice}
   \\
&2.& D_\mu \phi = ((D_\mu)_c e^{\frac{1}{2} \overleftarrow{\partial} \wedge
   P})(\phi_c e^{\frac{1}{2} \overleftarrow{\partial} \wedge P})
\end{eqnarray}
where
\begin{equation}
(D_\mu)_c = \partial_\mu + (A_\mu)_c
\end{equation}
and $(A_\mu)_c$ is the commutative gauge field, a function only of
$\hat{x}^c$. It is easy to see that the second choice does not
satisfy (\ref{commgauge}), but the first one does.The first choice is
also good because it preserves statistics, Poincar\'{e} and gauge
invariance.

As regards gauge invariance, we can see it as follows. The generators
of gauge transformations are the same as those for
$\theta^{\mu\nu}=0$. If we consider $D_\mu \phi |0\rangle$, it is the
same as the action of $(D_\mu)_c\phi_c$ on the Fock vacuum. Hence it
transforms correctly.

To see the compatibility of gauge transformations and statistics, let
us look at the operator product $(D_\mu \phi)(x) (D_\nu \phi)(y)$ and
restrict it to the two-particle sector. It reads
\begin{equation}
(D_\mu)_c\phi_c (x) e^{-\frac{i}{2} \overleftarrow{\partial} \wedge
  \overrightarrow{\partial}} (D_\mu)_c \phi_c (y) )|0\rangle
  \label{gauge2part}
\end{equation}
The Gauss law operator only transforms the operator parts of $(D_\mu)_c
\phi_c$ which are the analogs of creation-annihilation operators
$a^\dagger(p), a(p)$. That is, if
\begin{equation}
(D_\mu)_c\phi_c (x) = \sum_n \alpha_\mu^n f_n(x),
\end{equation}
then Gauss law only transforms the operators $\alpha_\mu^n$. So under
gauge transformations $g$,
\begin{equation}
(D_\mu)_c\phi_c (x) e^{-\frac{i}{2} \overleftarrow{\partial} \wedge
  \overrightarrow{\partial}}  (D_\mu)_c \phi_c (y) |0\rangle
  \rightarrow (g \alpha^n_\mu)(g \alpha^m_\nu) (f_n(x)
  e^{-\frac{i}{2} \overleftarrow{\partial} \wedge
  \overrightarrow{\partial}} f_m(y)) |0\rangle
\end{equation}
Under the multiplication map, the exponential cancels out, and
\begin{equation}
m_\theta ((g \alpha^n_\mu)(g \alpha^m_\nu) (f_n(x)
  e^{-\frac{i}{2} \overleftarrow{\partial}_x \wedge
  \overrightarrow{\partial}_y} f_m(y))) |0\rangle = g
  [(D_\mu)_c\phi_c] g [(D_\nu)_c \phi_c] |0\rangle
\end{equation}

Note that since the symmetry generators are the same as those for
$\theta^{\mu\nu}=0$, so the $(F_{\mu \nu})^2$ term of the gauge field
interaction also transforms correctly. 

Any gauge group can be treated in this approach, unlike some other approaches.

Similar arguments can be made about the transformation properties
under the Poincar\'{e} group.

The relation between our covariant derivative and that of
\cite{aschieri} is similar to the relation between the corresponding
operators appropriate for diffeos. We previously described this
connection in Section 4. Thus the gauge covariant derivative $D_\mu$
of \cite{aschieri} acts with a $*$-product on fields. Consider a new
covariant derivative 
\begin{equation}
{\cal D}_\mu = D_\mu e^{-\frac{i}{2} ad
  \overleftarrow{\partial}_\lambda \theta^{\lambda\rho}
  \overrightarrow{\partial}_\rho} 
\end{equation}
acting with a $*$-product on fields $\alpha$. This action becomes our
action of $D_\mu$ on $\alpha$:
\begin{equation}
{\cal D}_\mu * \alpha = D_\mu \alpha.
\end{equation}
The relation between the two covariant derivatives can be understood
in this manner.

\section{Quantum Gauge Theory}

Having identified the correct covariant derivative, it is simple to
write down the Hamiltonian for gauge theories. The commutator of two
covariant derivatives gives us the curvature. Using (\ref{rightchoice})
\begin{eqnarray} 
[D_\mu, D_\nu] \phi &=& ([D_{\mu c}, D_{\nu,c}]\phi_c) e^{\frac{1}{2}
  \overleftarrow{\partial} \wedge P}, \\ 
&=& (F_{\mu\nu,c}\phi_c)e^{\frac{1}{2}\overleftarrow{\partial} \wedge P}
\end{eqnarray} 

As $F_{\mu\nu,c}$ above transforms covariantly under gauge
transformations, we can use it to construct the Hamiltonian for the
gauge theory. Thus pure gauge theories on the GM plane are identical
to their counterparts on commutative space. 

However, the coupling between matter and gauge field, which involves
the covariant derivative of the matter field, is different from its
commutative analog. As a result, the interaction Hamiltonian splits
into two parts:
\begin{equation}
H_\theta^{I} = \int d^3 x [{\cal H}_\theta^{MG} + {\cal H}_\theta^G],
  \quad  MG = {\rm matter-gauge}, \quad G = {\rm pure\,\, gauge\,\, field}
\end{equation}
\begin{eqnarray}
{\cal H}_\theta^{MG} &=& {\cal H}_0^{MG} e^{\frac{1}{2}
  \overleftarrow{\partial} \wedge P}, \label{mattergauge}\\
{\cal H}_\theta^G &=& {\cal H}_0^G \ .
\end{eqnarray}
We include matter-gauge field couplings and all matter couplings in
${\cal H}_\theta^{MG}$, while ${\cal H}_\theta^G$ contains only gauge
field terms.  For QED, ${\cal H}_\theta^G =0$, so as shown in
\cite{bpq}, the $S$-operator of the theory is the same as the
commutative case:
\begin{equation}
S_\theta^{QED} = S_0^{QED} \ .
\end{equation}
For the Standard Model (SM), ${\cal H}_\theta^G = {\cal H}_0^G \neq
0$. As this term has no statistics twist,
\begin{equation}
S_\theta^{SM} \neq S_0^{SM} \ .
\end{equation}
because of the cross-terms in the $S$-matrix between ${\cal
  H}_\theta^{MG}$ and ${\cal H}_\theta^G$. In particular, this inequality
  happens in QCD. Processes like $qg \rightarrow qg$ via a gluon
  exchange interaction actually also violate causality and Lorentz
  invariance, as we indicate below. \footnote{A more in-depth
  discussion of causality in noncommutative theories will be presented
  elsewhere \cite{bpqv2}.}

\begin{figure}
\centerline{\epsfig{figure=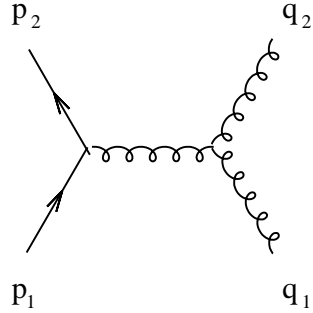,clip=4cm,width=4cm}}
\caption{A Feynman diagram with a non-trivial $\theta$-dependence}
\label{fig:qcd}
\end{figure}

The Feynman diagram responsible for this violation is shown in Fig
\ref{fig:qcd}. The twist of ${\cal H}_0^{MG}$ in (\ref{mattergauge})
changes the gluon propagator that connects the
quark-quark-gluon vertex to the 3-gluon vertex (and in fact to any
vertex containing just gluons). This propagator is
different from the usual one by its dependence on terms of the form
$\vec{\theta^0}.\vec{P}_{inc}$, where $(\vec{\theta^0})_i = \theta^{0i}$
and $\vec{P}_{inc}$ is the total momentum of the incoming
particles. Such dependence is clearly frame-dependent and violates
Lorentz invariance (Their $C$, $P$, and $T$ properties are discussed in
\cite{abjj}).  

{\bf Acknowledgments:} It is a pleasure to thank Earnest Akofor,
T. R. Govindarajan, Sang Jo and Anosh Joseph for discussions. APB
especially thanks Paolo Aschieri for a clarifying discussion about Eq
(7.13 -- 7.15). Some of the results of this papers overlap with some
of those of \cite{abjj}. The work of APB and BQ is supported in part
by DOE under grant number DE-FG02-85ER40231. The work of AP has been
supported by FAPESP grant number 06/56056-0.

\end{document}